# Inconsistencies in the current thermodynamic description of elastic solids

**József Garai[1] and Alexandre Laugier[2]**

[1] Department of Earth Sciences, University Park, PC 344, Miami, FL 33199, USA
[2] IdPCES - Centre d'affaires PATTON, 6, rue Franz Heller, bât. A, F35700 Rennes, France

E-mails:  [1] jozsef.garai@fiu.edu
          [2] alexandre.laugier@libertysurf.fr

**Abstract**

Using the contemporary thermodynamic equations of elastic solids leads to contradictions with the fundamental statements of thermodynamics. Two examples are presented to expose the inconsistencies. In example one the internal energy between the initial and final states shows path dependency while in example two changing the temperature of a system at constant volume produces mechanical work. These results are contradictory with the fundamentals of thermodynamics and indicate that the contemporary description of elastic solids needs to be revisited and revised.

In solid phase the relationship between the pressure and volume is described by the isothermal bulk modulus [$K_T$]:

$$K_T = -V\left(\frac{\partial p}{\partial V}\right)_T \tag{1}$$

It is assumed that the solid is homogeneous, isotropic, non-viscous and that the elasticity is linear. It is also assumed that the stresses are isotropic; therefore, the principal stresses can be identified as the pressure $p = \sigma_1 = \sigma_2 = \sigma_3$.

The function between temperature and volume is characterized by the volume coefficient of expansion [$\alpha_{V_p}$]:

$$\alpha_{V_p} = \frac{1}{V}\left(\frac{\partial V}{\partial T}\right)_p \tag{2}$$

The volume of a solid system at a given temperature and pressure can be calculated by employing the equations of the volume coefficient of expansion and the bulk modulus. Allowing one of the variables to change while the other one held constant the calculation can be done in two steps (Fig. 1),

$$(V_T)_{p=0} = V_0 e^{\int_{T=0}^{T} \alpha_{Vp=0} dT} \quad \text{or} \quad (V_p)_{T=0} = V_0 e^{\int_{p=0}^{p} -\frac{1}{K_{T=0}} dp} \quad (3)$$

and then

$$(V_T)_p = (V_p)_{T=0} e^{\int_{T=0}^{T} \alpha_{Vp} dT} \quad \text{or} \quad (V_p)_T = (V_T)_{p=0} e^{\int_{p=0}^{p} -\frac{1}{K_T} dp} \quad (4)$$

where $V_0$ is the volume at zero pressure and temperature. These two steps might be combined into one and the volume at a given p, and T can be calculated:

$$V_{p,T} = V_0 e^{\int_{T=0}^{T} \alpha_{Vp} dT - \int_{p=0}^{p} \frac{1}{K_T} dp} \quad (5)$$

According to the first law of thermodynamics the differentials of the internal energy [U] of a system can be calculated in the same way in solid phase as in gas phase thus the differential of the internal energy is equal with the sum of the differentials of the thermal and the mechanical energies.

$$dU = \delta q + \delta w = nc_V dT - pdV \quad (6)$$

where q is the heat, supplied to or taken away from the system, w is the mechanical work, done by or on the system, n is the number of moles, and $c_V$ is the molar heat capacity at constant volume. Heating up a system at zero pressure from zero temperature to temperature T the internal energy of the system will increase as:

$$(\Delta U_T)_{p=0} = \int_{T=0}^{T} nc_V dT \quad (7)$$

Applying pressure on the system at constant temperature and increasing the pressure from zero pressure to pressure p will change the internal energy of the system $(\Delta U_p)_T$ by the mechanical work $(\Delta w_p)_T$ done on the system. Besides the exponential nature of the isothermal curves (Eq. 3-4) the area (isothermal work function) under these curves almost identical with a triangle area. Small deviation could occur at extremely high pressures. The mechanical work, therefore, will be calculated as:

$$(\Delta U_p)_T = (\Delta w_p)_T = -\frac{1}{2} p (\Delta V_p)_T \tag{8}$$

where

$$(\Delta V_p)_T = V_0 e^{\int_{T=0}^{T} \alpha_{V_p} dT} \left( e^{\int_{p=0}^{p} -\frac{1}{K_T} dp} - 1 \right) \tag{9}$$

The internal energy of the system at temperature T and pressure p is

$$U_{p,T} = U_{p=0,T=0} + (\Delta U_T)_{p=0} + (\Delta U_p)_T = U_{p=0,T=0} + \int_{T=0}^{T} nc_V dT - \frac{1}{2} pV_0 e^{\int_{T=0}^{T} \alpha_{V_p} dT} \left( e^{\int_{p=0}^{p} -\frac{1}{K_T} dp} - 1 \right) \tag{10}$$

Using subscript i and f for the initial and final state respectively and assuming that

$$U_{p=0,T=0} = U_i \tag{11}$$

and

$$U_{p,T} = U_f \tag{12}$$

then the internal energy difference between the initial and final state is:

$$\Delta U_{path-1} = U_f - U_i = \int_{T=0}^{T} nc_V dT - \frac{1}{2} pV_0 e^{\int_{T=0}^{T} \alpha_{V_p} dT} \left( e^{\int_{p=0}^{p} -\frac{1}{K_T} dp} - 1 \right). \tag{13}$$

Subscript path-1 indicates the path which was followed between the initial and final conditions. The internal energy is state function; therefore, the internal energy change between the initial and final states is the same no matter what path has been followed between these two states (Figure 2a). Using the previous example the internal energy of the final state should be the same if the pressure is applied on the system at zero temperature first and then the system is heated up to temperature T at constant pressure (path-2).

$$U_{f,path-1} = U_{f,path-2} \iff U_{f,path-1} - U_i = U_{f,path-2} - U_i \iff \Delta U_{path-1} = \Delta U_{path-2} \tag{14}$$

At zero temperature the pressure increase from zero to p will do work on the system.

$$(\Delta U_p)_{T=0} = (\Delta w_p)_{T=0} = -\frac{1}{2} p (\Delta V_p)_{T=0} \tag{15}$$

where

$$(\Delta V_p)_{T=0} = V_0 \left( e^{\int_{p=0}^{p} -\frac{1}{K_T} dp} - 1 \right) \tag{16}$$

Increasing the temperature at constant pressure from zero temperature to temperature T involves both thermal $(\Delta q_T)_p$ and mechanical energies $(\Delta w_T)_p$.

$$(\Delta U_T)_p = (\Delta q_T)_p + (\Delta w_T)_p \tag{17}$$

These energies can be calculated as:

$$(\Delta q_T)_p = \int_{T=0}^{T} nc_V dT \quad \text{and} \quad (\Delta w_T)_p = -p(\Delta V_T)_p \tag{18}$$

where

$$(\Delta V_T)_p = V_0 e^{\int_{p=0}^{p} -\frac{1}{K_T} dp} \left( e^{\int_{T=0}^{T} \alpha_{V_p} dT} - 1 \right). \tag{19}$$

Summing the energies between the initial and the final state gives the total internal energy change for path-2.

$$\Delta U_{\text{path}-2} = U_f - U_i = -\frac{1}{2} p(\Delta V_p)_{T=0} + \int_{T=0}^{T} nc_V dT - p(\Delta V_T)_p \tag{20}$$

Using the internal energy equivalency for the two different paths and deducting the internal energy change of path one from path two gives:

$$\frac{1}{2} p\left[(\Delta V_p)_{T=0} - (\Delta V_p)_T\right] + p(\Delta V_T)_p = 0. \tag{21}$$

After simplifications it can be written:

$$\frac{1}{2} pV_0 \left( e^{\int_{T=0}^{T} \alpha_{V_p} dT} - 1 \right) \left( 1 + e^{\int_{p=0}^{p} -\frac{1}{K_T} dp} \right) = 0. \tag{22}$$

At pressures higher than zero the equality implies that either the factor

$$\left( e^{\int_{T=0}^{T} \alpha_{V_p} dT} - 1 \right) = 0 \quad \text{or} \quad \left( 1 + e^{\int_{p=0}^{p} -\frac{1}{K_T} dp} \right) = 0. \tag{23}$$

Since

$$e^{\int_{p=0}^{p} -\frac{1}{K_T} dp} > 0 \tag{24}$$

then

$$\left( 1 + e^{\int_{p=0}^{p} -\frac{1}{K_T} dp} \right) \neq 0 \tag{25}$$

Assuming positive value for the volume coefficient of expansion at temperatures higher than zero

$$e^{\int_{T=0}^{T} \alpha_{V_p} dT} > 1 \tag{26}$$

results that

$$\left(e^{\int_{T=0}^{T}\alpha_{V_p}dT}-1\right)\neq 0 \tag{27}$$

It can be concluded that the equality $\Delta U_{path-1} = \Delta U_{path-2}$ is incorrect in this framework.

Another discrepancy in the current description of elastic solids can be noted by heating up a system at constant volume. The internal energies for the initial and final states are

$$U_i = q_i + w_i \quad \text{and} \quad (U_f)_V = q_i + w_i + \int_{T_i}^{T_f} nc_V dT = q_i + \int_{T_i}^{T_f} nc_V dT + w_f. \tag{28}$$

Keeping the volume constant resulting that the initial and final volumes are the same.

$$V_i = V_f \quad \text{or} \quad \frac{V_f}{V_i} = 1 \tag{29}$$

Using equation 5 it can be seen that the volume remains constant if the equality

$$e^{\int_{T_i}^{T_f}\alpha_{V_p}dT} e^{\int_{p_i}^{p_f}-\frac{1}{K_T}dp} = 1 \tag{30}$$

is satisfied. The mechanical work, stored by the system in the initial state is

$$w_i = -\frac{1}{2}p_i V_0 e^{\int_{T=0}^{T_i}\alpha_{V_p}dT}\left(e^{\int_{p=0}^{p_i}-\frac{1}{K_T}dp}-1\right). \tag{31}$$

The work in the final state can be written as:

$$w_f = -\frac{1}{2}p_f V_0 e^{\int_{T=0}^{T_f}\alpha_{V_p}dT}\left(e^{\int_{p=0}^{p_f}-\frac{1}{K_T}dp}-1\right). \tag{32}$$

The mechanical work of the final state (using Eq. 32) can be factorized as:

$$w_f = -\frac{1}{2}p_f V_0 e^{\int_{T=0}^{T_i}\alpha_{V_p}dT} e^{\int_{T_i}^{T_f}\alpha_{V_p}dT}\left(e^{\int_{p=0}^{p_i}-\frac{1}{K_T}dp} e^{\int_{p_i}^{p_f}-\frac{1}{K_T}dp}-1\right) \tag{33}$$

and

$$w_f = -\frac{1}{2}p_f V_0 e^{\int_{T=0}^{T_i}\alpha_{V_p}dT}\left(e^{\int_{p=0}^{p_i}-\frac{1}{K_T}dp}-e^{\int_{T_i}^{T_f}\alpha_{V_p}dT}\right). \tag{34}$$

According to the first law of thermodynamics when a system is heated up constant volume mechanical work is not done on or by the system. The mechanical energy of the system in the initial and final state should be the same.

$$(w_i)_V = (w_f)_V \tag{35}$$

Using the equality of the initial and final mechanical work it can be written

$$p_i \left( e^{\int_{p=0}^{p_i} -\frac{1}{K_T} dp} - 1 \right) = p_f \left( e^{\int_{p=0}^{p_i} -\frac{1}{K_T} dp} - e^{\int_{T_i}^{T_f} \alpha_{V_p} dT} \right). \tag{36}$$

and then

$$e^{\int_{p=0}^{p_i} -\frac{1}{K_T} dp} (p_f - p_i) + p_i = p_f e^{\int_{T_i}^{T_f} \alpha_{V_p} dT} . \tag{37}$$

Since

$$p_f = (p_f - p_i) + p_i \tag{38}$$

then

$$(p_f - p_i) e^{\int_{p=0}^{p_i} -\frac{1}{K_T} dp} + p_i = (p_f - p_i) e^{\int_{T_i}^{T_f} \alpha_{V_p} dT} + p_i e^{\int_{T_i}^{T_f} \alpha_{V_p} dT} \tag{39}$$

Assuming that both the volume coefficient of expansion and the bulk modulus are positive and that the final temperature is higher than the initial one and the initial pressure is nonzero then

$$e^{\int_{p=0}^{p_i} -\frac{1}{K_T} dp} < 1 < e^{\int_{T_i}^{T_f} \alpha_{V_p} dT} \tag{40}$$

resulting that

$$(p_f - p_i) e^{\int_{p=0}^{p_i} -\frac{1}{K_T} dp} + p_i < (p_f - p_i) e^{\int_{T_i}^{T_f} \alpha_{V_p} dT} + p_i e^{\int_{T_i}^{T_f} \alpha_{V_p} dT} \tag{41}$$

This result contradicts with the original assumption $w_i = w_f$ indicating that work had been done at constant volume. The inequality of the initial and final work can be easily justified by the visual inspection of Figure 2b.

In both examples the currently accepted thermodynamic description of elastic solids gives erroneous results. These erroneous results indicate that either the original assumption, the internal energy is a state function, or the thermodynamic equations used for the calculations are incorrect. It is very unlikely the state function of the internal energy can be questioned leaving no other chose then assuming that the inconsistencies in the thermodynamics of elastic solids arise from incorrect thermodynamics equations.

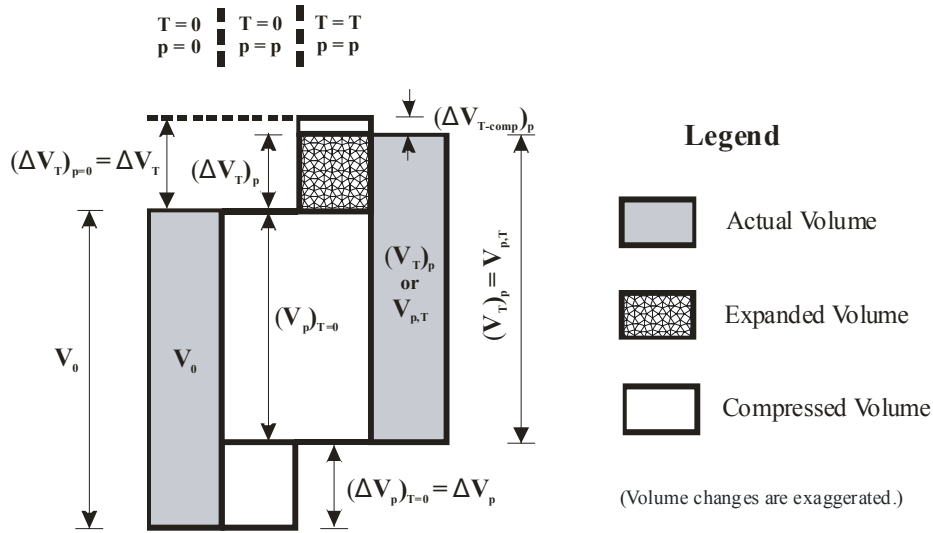

**Figure 1**. Schematic figure of the volume changes caused by the pressure and the temperature. The analytical description of the different volumes is given in the text.

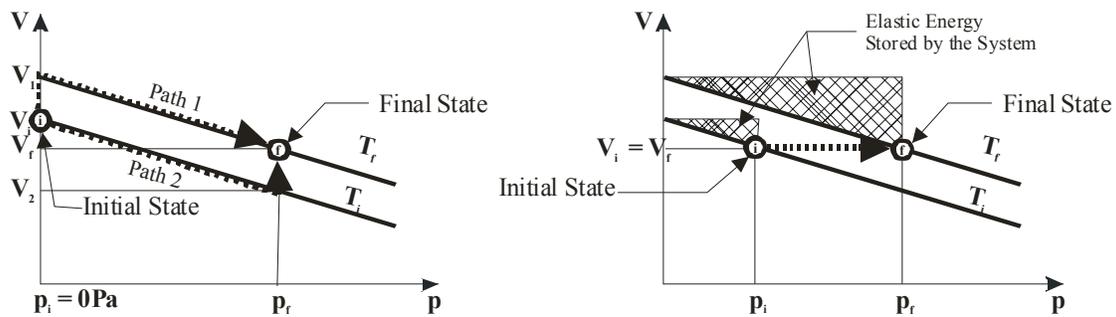

**Figure 2.** Examples for the existing inconsistencies in the current thermodynamic description of elastic solids. (a) The internal energy between an initial and final state shows path dependency if the current conventional thermodynamic expressions are used. (b) Heating up a system at constant volume changes the mechanical work stored by the system. According to the current thermodynamic establishment no mechanical work is done by or on the system if the volume is held constant.